\documentclass[aps,pre,showpacs,twocolumn]{revtex4}
\usepackage{bm}
\usepackage{graphicx}
\bibstyle{approve.bib}
\usepackage{amssymb}
\usepackage{epstopdf}
\usepackage{color}
\newcommand{\be}{\begin{equation}}
\newcommand{\ee}{\end{equation}}
\newcommand{\bea}{\begin{eqnarray}}
\newcommand{\eea}{\end{eqnarray}}

\newcommand{\br}{\mathbf{r}}

\newcommand{\e}{\varepsilon}

\begin{document}

\title{Ionic current inversion in pressure-driven polymer translocation through nanopores}

\author{Sahin Buyukdagli$^{1,2}$\footnote{email:~\texttt{Buyukdagli@fen.bilkent.edu.tr}}, Ralf Blossey$^{2}$\footnote{email:~\texttt{Ralf.Blossey@iri.univ-lille1.fr}},  and T. Ala-Nissila$^{3,4}$\footnote{email:~\texttt{Tapio.Ala-Nissila@aalto.fi}}}
\affiliation{$^{1}$Department of Physics, Bilkent University, Ankara 06800, Turkey\\
$^{2}$Institut de Recherche Interdisciplinaire USR3078 CNRS and Universit\'e Lille I, Parc de la Haute Borne, 52 Avenue de Halley, 59658 Villeneuve d'Ascq, France\\
$^{3}$Department of Applied Physics and COMP Center of Excellence, Aalto University School of Science, P.O. Box 11000, FI-00076 Aalto, Espoo, Finland\\
$^{4}$Department of Physics, Brown University, Providence, Box 1843, RI 02912-1843, U.S.A.}
\date{\today}

\begin{abstract}
We predict streaming current inversion with multivalent counterions in hydrodynamically driven polymer translocation events from a correlation-corrected charge transport theory including charge fluctuations around mean-field
electrostatics. In the presence of multivalent counterions, electrostatic many-body effects result in the reversal of the DNA charge. The attraction of anions to the charge-inverted DNA molecule reverses the sign of the ionic current
through the pore. Our theory allows for a comprehensive understanding of the complex features of the resulting streaming currents. The underlying mechanism is an efficient way to detect DNA charge reversal in
pressure-driven translocation experiments with multivalent cations.
\end{abstract}

\date{\today}
\maketitle

Coulombic interactions play a fundamental role in biological systems as well as in various nanotechnologies currently under development, from biopolymer analysis~\cite{e1,Huang2010} to nanofluidic transport~\cite{Miles2001,Gu2000,Stein2006}. Among them, the controlled translocation of biopolymers thorough nanopores has witnessed a rapid advancement recently~\cite{e2,e3,e4,e5,e6,e7,e8,e9,e10,e11,e12}. Polymer
translocation aims at probing the biopolymer sequence via the characteristics of the ionic current thorough the pore. The improvement of  this method requires an accurate understanding  of the electrohydrodynamics at play.
This challenge has not been fully met since previous theoretical approaches either focused exclusively on entropic effects~\cite{n1,n2,n3,n4,n5} or made use of mean-field (MF) electrostatics known to be inaccurate with multivalent ions~\cite{Ghosal2006,Ghosal2007,Keijan2009}. We have recently taken a step forward and shown that in \textit{electrophoretic} DNA translocation, \textcolor{black}{DNA charge inversion induced by multivalent ions reverses the direction of the polymer without affecting the sign of the electrophoretic current~\cite{Buyuk2014II}. This result indicates the possibility to use the charge correlation effect in order to control the DNA translocation speed whose minimization is a required condition to improve the resolution of this method.}  In the present letter, we focus on \textit{pressure-driven} translocation and show that the presence of multivalent ions in the solution results in the inversion of the streaming current thorough the pore, \textcolor{black}{an effect absent in electrophoretic polymer transport}. Our prediction is supported by previous nanofluidics experiments where ionic current inversion had been observed in nanoslits confining multivalent charges~\cite{Heyden2006}.

\begin{figure}
\includegraphics[width=1\linewidth]{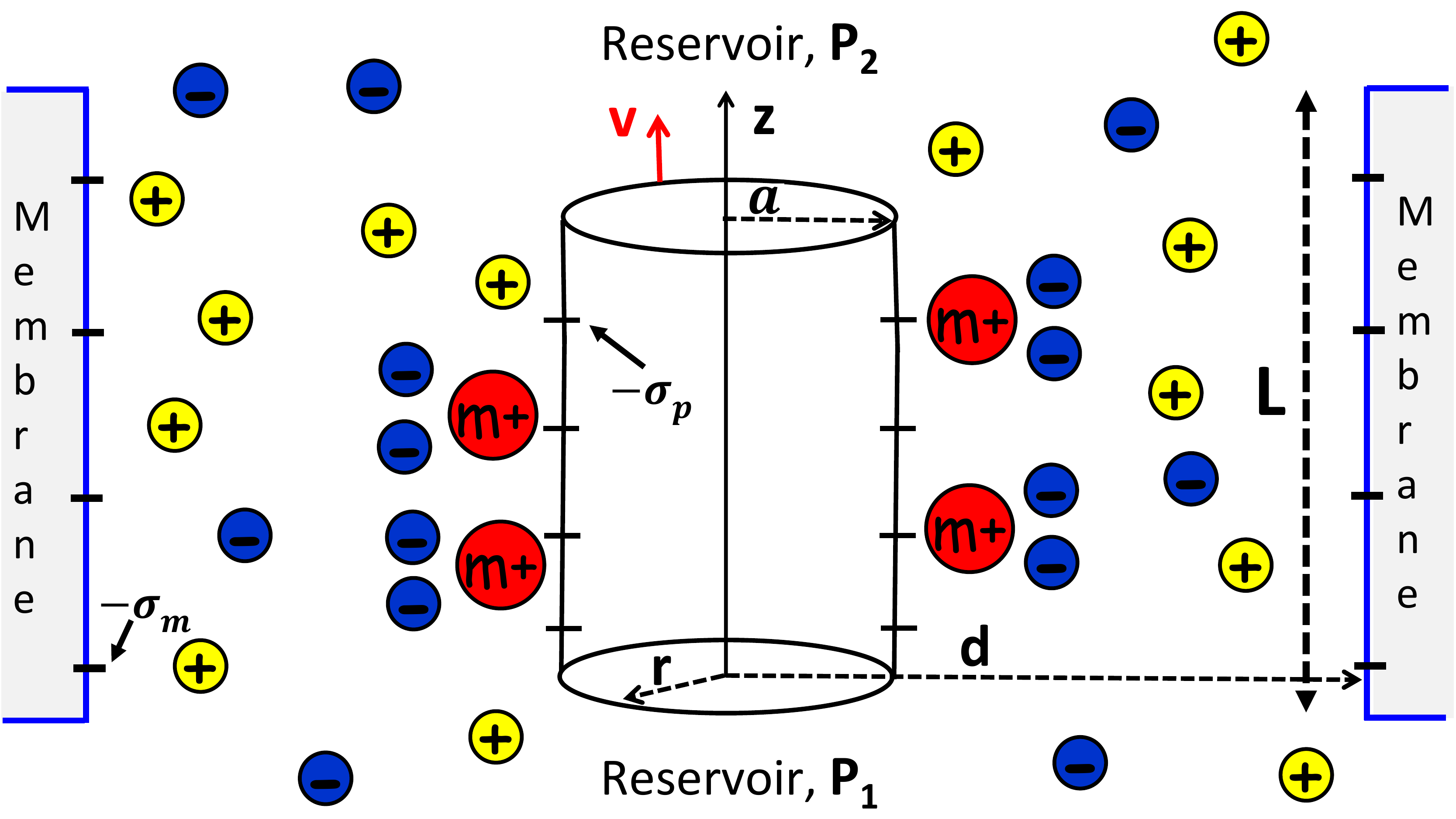}
\caption{(Color online) Schematic representation of the nanopore. The cylindrical polyelectrolyte of radius $a=1$ nm and surface charge $-\sigma_p$ is confined to the cylindrical pore of radius $d=3$ nm, length $L=340$ nm, and wall charge $-\sigma_m$. The pore connects the reservoirs at the hydrodynamic pressures $P_1$ and $P_2$. Black circles denote the multivalent cations $\mbox{I}^{m+}$ contained in the electrolyte mixture $\mbox{KCl}+\mbox{ICl}_m$.}
\label{Fig1}
\end{figure}

Our translocating polymer model system depicted in Fig.~\ref{Fig1} consists of a charged liquid confined between the cylindrical polymer and pore with negative smeared charge distributions $-\sigma_m$ and $-\sigma_p$. The pore radius $d=3$ nm and length $L=340$ nm lie in the typical range of solid-state nanopores~\cite{e9}.  The polyelectrolyte radius is taken as the radius of double-stranded (ds)-DNA molecules $a=1$ nm~\cite{Buyuk2014II}.  Driven by the pressure gradient $\Delta P_z=P_2-P_1>0$ at the pore edges, the polyelectrolyte translocates along the $z$-axis. We neglect any off-axis fluctuations. Moreover, we have recently found that dielectric discontinuities play a minor role in solid-state pores with radius beyond the nanometer scale~\cite{Buyuk2014II}. Thus, we assume that the whole system has the dielectric permittivity of water, $\e_w=80$.  The ionic current through the pore is given by the number of flowing charges per unit time integrated over the cross-section of the channel,
\be\label{str1}
\rm{I}_{\rm{str}}=2\pi e\int_{a^*}^{d^*}\mathrm{d}rr\rho_c(r)u(r).
\ee
In Eq.~(\ref{str1}), $e$ is the elementary charge, $u(r)$ the liquid velocity,  and  $\rho_c(r)$ the ionic charge density. Because the length of the pore is much larger than its radius $L\gg d$, the liquid velocity and density are assumed to depend solely on the radial distance. Furthermore, by introducing the effective polymer radius $a^*=a+a_{st}$ and pore radius $d^*=d-a_{st}$ where the Stern layer $a_{st}=2$ {\AA} corresponds to the characteristic hydration radius of multivalent cations~\cite{Ohtaki1993,rem0}, we account for the stagnant ion layer close to the charged nanopore and polyelectrolyte surfaces~\cite{Joly2004,Qiao2004}.

The correlation-corrected charge density in Eq.~(\ref{str1}) is computed within the one-loop theory of electrolyte mixtures in cylindrical pores~\cite{Buyuk2014},
\be\label{chden}
\rho_c(r)=\sum_iq_in_i(r)\left[1-q_i\phi_1(r)-\frac{q_i^2}{2} \delta v(r)\right],
\ee
with the MF-level ionic number density
\be\label{aux1}
n_i(r)=\rho_{ib}e^{-q_i\phi_0(r)}\theta(r-a)\theta(d-r).
\ee
The summation in Eq.~(\ref{chden}) runs over the ionic species in the solution, with each species $i$ of valency $q_i$ and reservoir concentration $\rho_{ib}$. The external potential $\phi_0(r)$ determining in Eq.~(\ref{aux1}) the MF-level ion densities follows from the solution of the Poisson-Boltzmann (PB) equation
\bea\label{eqt1}
\nabla^2\phi_0(r)+4\pi\ell_B\sum_iq_in_i(r)=-4\pi\ell_B\sigma(r),
\eea
with the pore and polymer charge distribution function $\sigma(r)=-\sigma_m\delta(r-d)-\sigma_p\delta(r-a)$. The one-loop potential correction $\phi_1(r)$ and the ionic self-energy $\delta v(r)$ including quadratic fluctuations around the MF potential are obtained from the relations
\bea
\label{eqt2}
&&\phi_1(r)=-\frac{1}{2}\sum_iq_i^3\int\mathrm{d}\br'v(\br,\br')n_i(r')\delta v(r')\\
\label{self}
&&\delta v(r)= \lim_{\br'\to\br}\left\{v(\br,\br')-v_c^b(\br-\br')+\ell_B\kappa_b\right\}.
\eea
The electrostatic propagator $v(\br,\br')$ in Eqs.~(\ref{eqt2})-(\ref{self}) is the solution of the kernel equation~\cite{Buyuk2014}
\bea\label{eqt3}
&&\nabla^2 v(\br,\br')-4\pi\ell_B\sum_iq_i^2n_i(r)v(\br,\br')=-4\pi\ell_B\delta(\br-\br'),\nonumber\\
\eea
the Bjerrum length in water at temperature $T=300$ K is $\ell_B=e^2/(4\pi\e_w k_BT)\simeq 7$ {\AA}, the Debye-H\"uckel screening parameter reads as
$\kappa_b^2=4\pi\ell_B\sum_iq_i^2\rho_{ib}$, and the Coulomb potential in an ion-free bulk solvent is given by $v_c^b(r)=\ell_B/r$.

\begin{figure}
\includegraphics[width=1\linewidth]{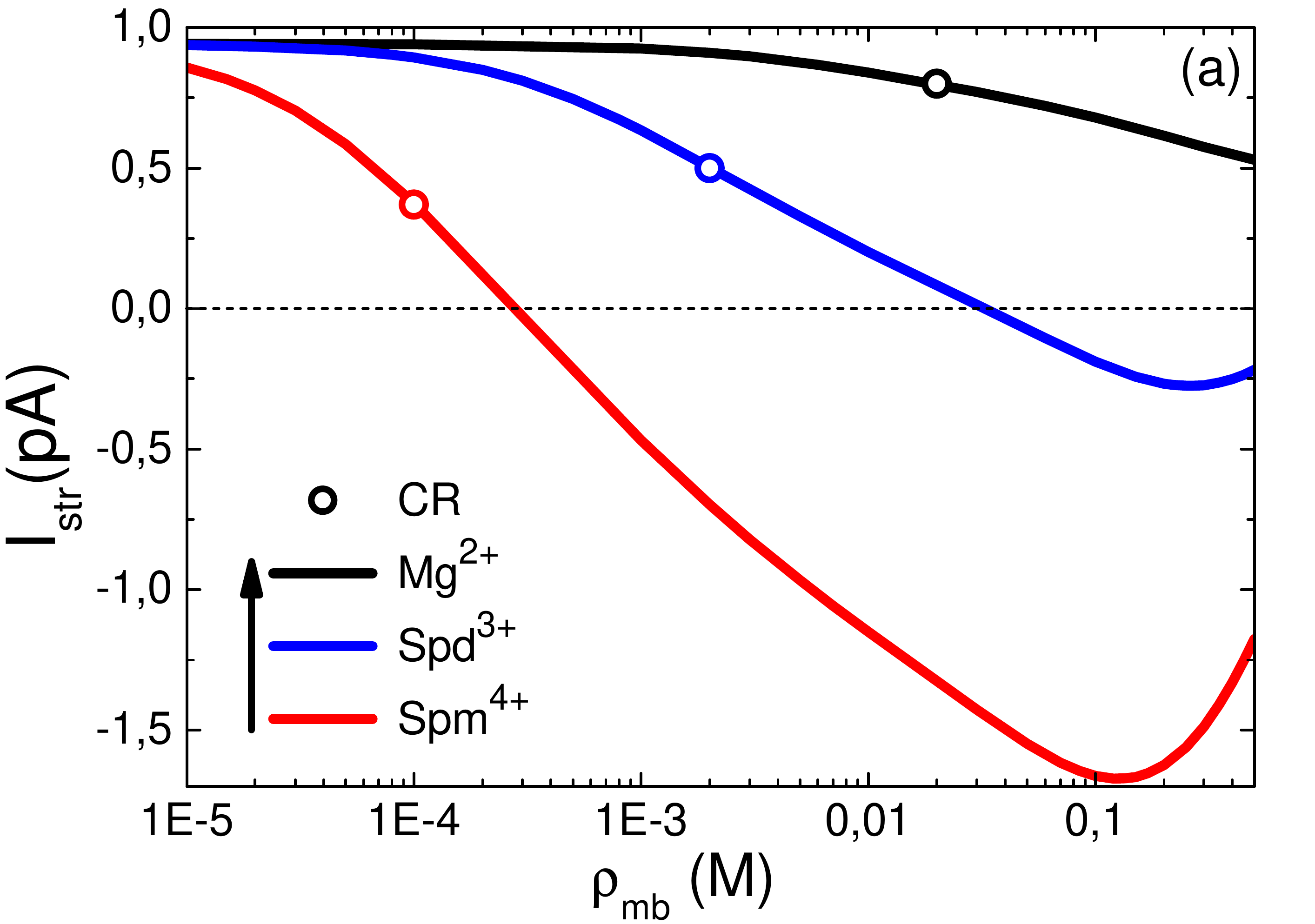}
\includegraphics[width=1\linewidth]{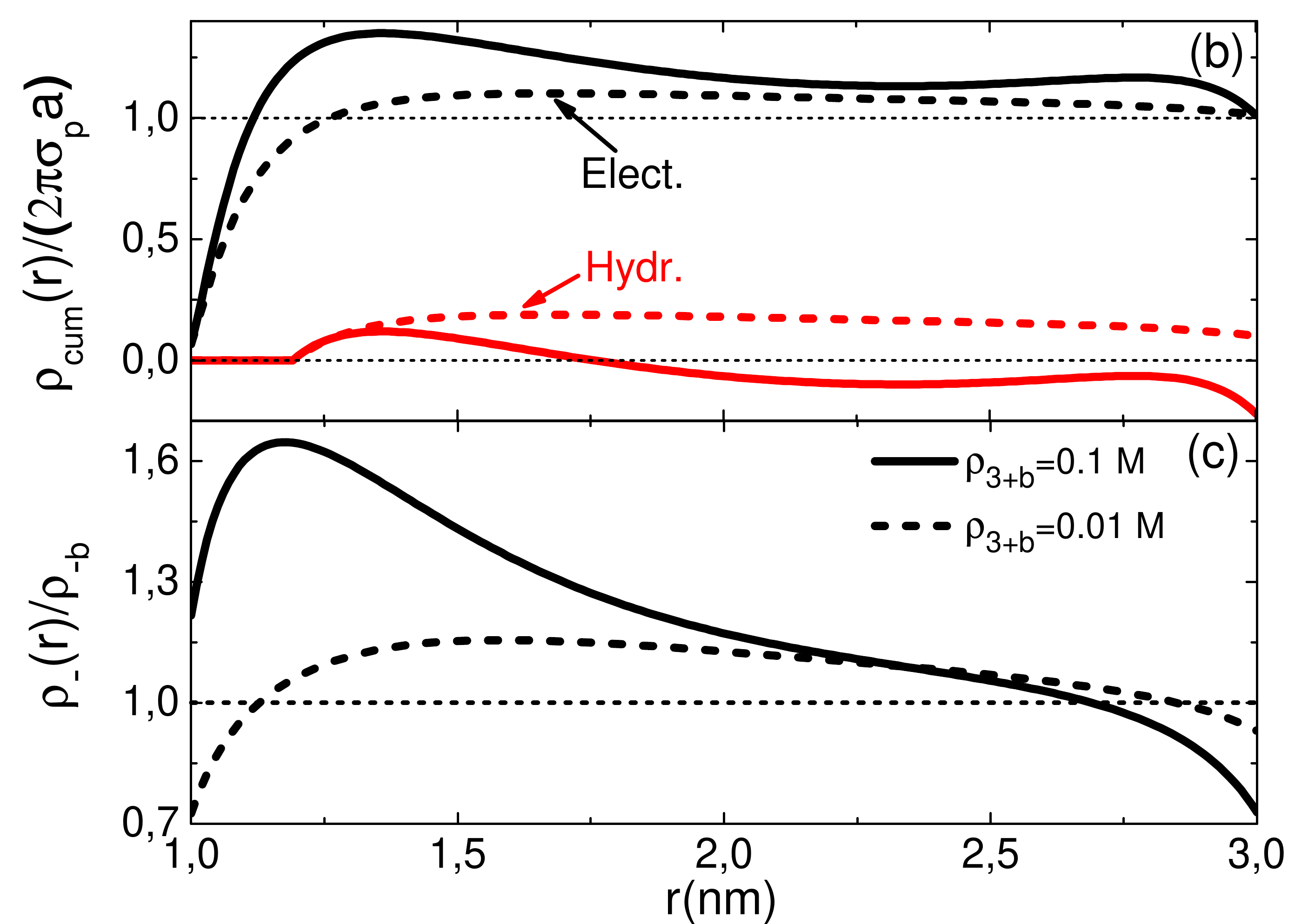}
\caption{(Color online) (a) Streaming current curves at the pressure gradient $\Delta P_z=1$ bar against the reservoir density of the multivalent counterion species given in the legend. Open circles mark the charge reversal (CR) points. (b) Electrostatic (black curves) and hydrodynamic (black curves) cumulative charge densities, and (c) $\mbox{Cl}^-$ densities in the  $\mbox{KCl}+\mbox{SpdCl}_3$ liquid at the reservoir concentrations $\rho_{3+b}=0.01$ M (dashed curves) and $0.1$ M (solid curves). The neutral nanopore ($\sigma_m=0$)   contains a ds-DNA of charge density $\sigma_p=0.4$ $e/\mbox{nm}^2$, with the bulk $\mbox{K}^+$ density $\rho_{+b}=0.1$ M in all figures. }
\label{Fig2}
\end{figure}
The liquid velocity $u(r)$ in Eq.~(\ref{str1}) satisfies the Stokes equation with applied pressure
\be\label{s1}
\eta\textcolor{black}{\nabla^2 u(r)}+\frac{\Delta P_z}{L}=0,
\ee
with the viscosity coefficient of water $\eta=8.91\times 10^{-4}\;\mathrm{Pa}\;\mathrm{s}$. Solving Eq.~(\ref{s1}) in the pore with the hydrodynamic boundary conditions $u(d^*)=0$ and $u(a^*)=v_T$ where $v_T$ stands for the translocation velocity of the polymer, and taking into account that the viscous friction force $F_v=2\pi a^*\eta u'(a^*)$ vanishes in the stationary state regime, the streaming current velocity follows in the form of a generalized Poisseuille profile,
\be\label{po}
u(r)=\frac{\Delta P_z}{4\eta L}\left[{d^*}^2-r^2+2{a^*}^2\ln\left(\frac{r}{d^*}\right)\right].
\ee
The charge density~(\ref{chden}) and the velocity profile~(\ref{po}) complete the calculation of the streaming current in Eq.~(\ref{str1}).

We consider first a ds-DNA with charge density $\sigma_p=0.4$ $e/\mbox{nm}^2$ confined to a neutral pore  ($\sigma_m=0$)~\cite{rem1}. Fig.~\ref{Fig2}(a) displays the streaming current of the electrolyte mixture $\mbox{KCl}+\mbox{ICl}_m$ against the reservoir density of the multivalent cation species $\mbox{I}^{m+}$ specified in the legend. One sees that for all multivalent counterion species, the ionic current is positive at dilute concentrations. This limit corresponds qualitatively to the MF-transport regime driven by the attraction of cations into the pore by the negative charge of the translocating DNA. With the increase of the magnesium density in the  $\mbox{KCl}+\mbox{MgCl}_2$ liquid, the ionic current diminishes but stays positive. Increasing the multivalent ion density in the electrolyte mixtures with $\mbox{Spd}^{3+}$ or $\mbox{Spm}^{4+}$ molecules, the ionic current first vanishes and then becomes negative above the respective reservoir concentrations $\rho_{3+b}\approx3\times10^{-2}$ M and $\rho_{4+b}\approx3\times10^{-4}$ M.  In the density regime $\rho_{mb}\gtrsim0.1-0.2$ M, the negative currents reach a minimum and start rising. We emphasize that this behaviour qualitatively agrees with the trend of the streaming currents measured in nanofluidic experiments with trivalent cations~\cite{Heyden2006} and investigated within numerical approaches~\cite{Labbez2009,Hoffmann2013}.

The negative sign of the ionic current indicates a net negative charge density between the pore and the DNA molecule, an effect that cannot be explained by MF arguments. In order to scrutinize the physical mechanism behind the sign reversal of the ionic current, we display in Fig.~\ref{Fig2}(b) the electrostatic cumulative charge density $\rho_{\rm{cum}}(r)=2\pi\int_a^r\mathrm{d}r'r'\rho_{c}(r')$ and the hydrodynamic cumulative charge density  $\rho^*_{\rm{cum}}(r)=2\pi\int_{a^*}^r\mathrm{d}r'r'\rho_{c}(r')$ of the  $\mbox{KCl}+\mbox{SpdCl}_3$ liquid rescaled with the DNA charge. The hydrodynamic cumulative density accounts exclusively for the charges contributing to the streaming flow. We also report in Fig.~\ref{Fig2}(c) the local densities of $\mbox{Cl}^-$ ions. At the bulk concentration $\rho_{3+b}=0.01$ M, the net cumulative charge density is seen to exceed the DNA charge at $r\gtrsim1.25$ nm. This is the signature of the DNA charge reversal induced by pronounced electrostatic correlations between $\mbox{Spd}^{3+}$ counterions bound to DNA. One notes that as a result of the charge-reversal effect, a weak $\mbox{Cl}^-$ excess $\rho_{-}(r)>\rho_{-b}$ takes place between the nanopore and the DNA molecule. However, because the anion attraction to the charge-inverted DNA is not significant,  the hydrodynamic flow charge is dominated by counterions and stays positive at the corresponding $\mbox{Spd}^{3+}$ density, i.e. $\rho_{\rm{cum}}(r)>0$ for $a^*<r<d$. By increasing the bulk spermidine concentration to $\rho_{3+b}=0.1$ M where one gets into the negative current regime in Fig.~\ref{Fig2}(a), the intensification of the DNA charge reversal (see Fig.~\ref{Fig2}(b)) leads to an amplified $\mbox{Cl}^-$ adsorption into the pore (see Fig.~\ref{Fig2}(c)). This significant anion excess results in a negative hydrodynamic charge density $\rho_{\rm{cum}}(d)<0$ leading to the negative ionic current thorough the pore. We also found that rising the spermidine concentration beyond $\rho_{3+b}\approx0.2$ M, the high anion concentration in the pore starts reducing the charge-reversal effect driving the ionic current inversion. This explains the minimum of the streaming current curves in Fig.~\ref{Fig2}(a).
\begin{figure}
\includegraphics[width=1.\linewidth]{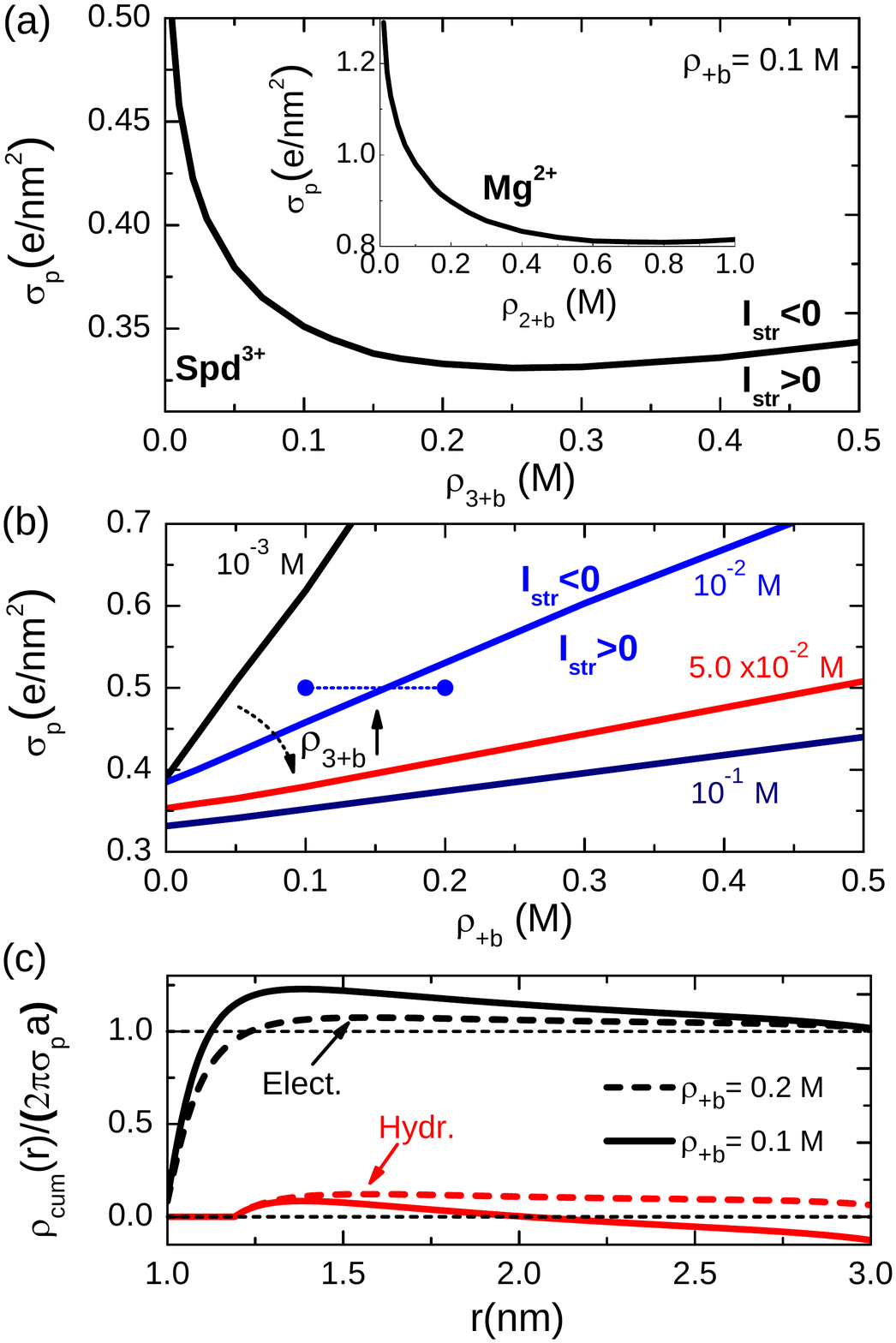}
\caption{(Color online) Characteristic polyelectrolyte charge against (a) $\mbox{Spd}^{3+}$ concentration ($\mbox{K}^+$ density fixed at $0.1$ M) and (b) $\mbox{K}^+$  concentration curves ($\mbox{Spd}^{3+}$ densities displayed next to each curve) separating phase domains with positive current $\mbox{I}_{\rm{str}}>0$ (below the curves) and negative current $\mbox{I}_{\rm{str}}<0$ (above the curves) in the  $\mbox{KCl}+\mbox{SpdCl}_3$ liquid. The inset in (a) displays the phase diagram of the main plot for the $\mbox{KCl}+\mbox{MgCl}_2$ liquid. (c) \textcolor{black}{ Electrostatic (black curves) and hydrodynamic (black curves) cumulative charge densities at the bulk $\mbox{K}^+$ concentration $\rho_{+b}=0.1$ M (solid curves) and $\rho_{+b}=0.2$ M (dashed curves). Polymer charge and bulk $\mbox{Spd}^{3+}$ densities are $\sigma_p=0.5$ $e/\mbox{nm}^2$ and $\rho_{3+b}=0.01$ M.} The remaining parameters are the same as in Fig.~\ref{Fig2}.}
\label{Fig3}
\end{figure}

These results show that although the DNA charge reversal is a necessary condition for the occurrence of the current inversion, the charge-reversal effect has to be strong enough for the adsorbed anions to compensate the contribution from cations to the streaming current. In particular with $\mbox{Mg}^{2+}$ ions in contact with the ds-DNA molecule, the charge reversal that takes place at $\rho_{2+b}\approx2.0\times10^{-2}$ M remains insufficient to invert the streaming current up to the highest bulk density $\rho_{2+b}=0.5$ M considered in Fig.~\ref{Fig2}(a). In the electrolyte mixtures with spermidine and spermine molecules, the charge reversal densities $\rho_{3+b}\approx2.0\times10^{-3}$ M  and $\rho_{4+b}\approx10^{-4}$ M are lower than the current inversion densities by several factors. These observations contradict the conclusion of Ref.~\cite{Heyden2006} where the authors had argued a one-to-one mapping between the reversal of the membrane charge and the sign reversal of the streaming current. In order to determine the charge densities where the current inversion effect is expected in translocation experiments, we plotted in Fig.~\ref{Fig3}(a) the critical polymer charge versus multivalent ion density curves where the streaming current switches from positive to negative. The main plot shows that in the $\mbox{KCl}+\mbox{SpdCl}_3$ liquid, the critical polymer charge for current inversion drops with increasing spermidine density  ($\rho_{3+b}\uparrow\sigma_p^*\downarrow$) up to  the point $\rho_{3+b}\approx0.2$ M where it reaches a minimum and rises beyond this value ($\rho_{3+b}\uparrow\sigma_p^*\uparrow$). The minimum of this curve at $\sigma_p\approx 0.33$ $e/\mbox{nm}^2$ corresponds to the lowest polymer charge below which it is impossible to observe ionic current inversion regardless of the bulk spermidine concentration. How is the phase diagram modified if one replaces the spermidine molecules with magnesium ions? The inset of Fig.~\ref{Fig3}(a) indicates that in the $\mbox{KCl}+\mbox{MgCl}_2$ liquid, the critical polymer charges for current inversion are about three times higher than in the $\mbox{KCl}+\mbox{SpdCl}_3$ electrolyte.  The main prediction of this diagram is that with $\mbox{Mg}^{2+}$ cations, it is impossible to observe the sign reversal of the streaming current with translocating ds-DNA molecules whose smeared charge density $\sigma_p=0.4$ $e/\mbox{nm}^2$ is located well below the minimum of the current inversion curve.

\begin{figure}
\includegraphics[width=1.6\linewidth]{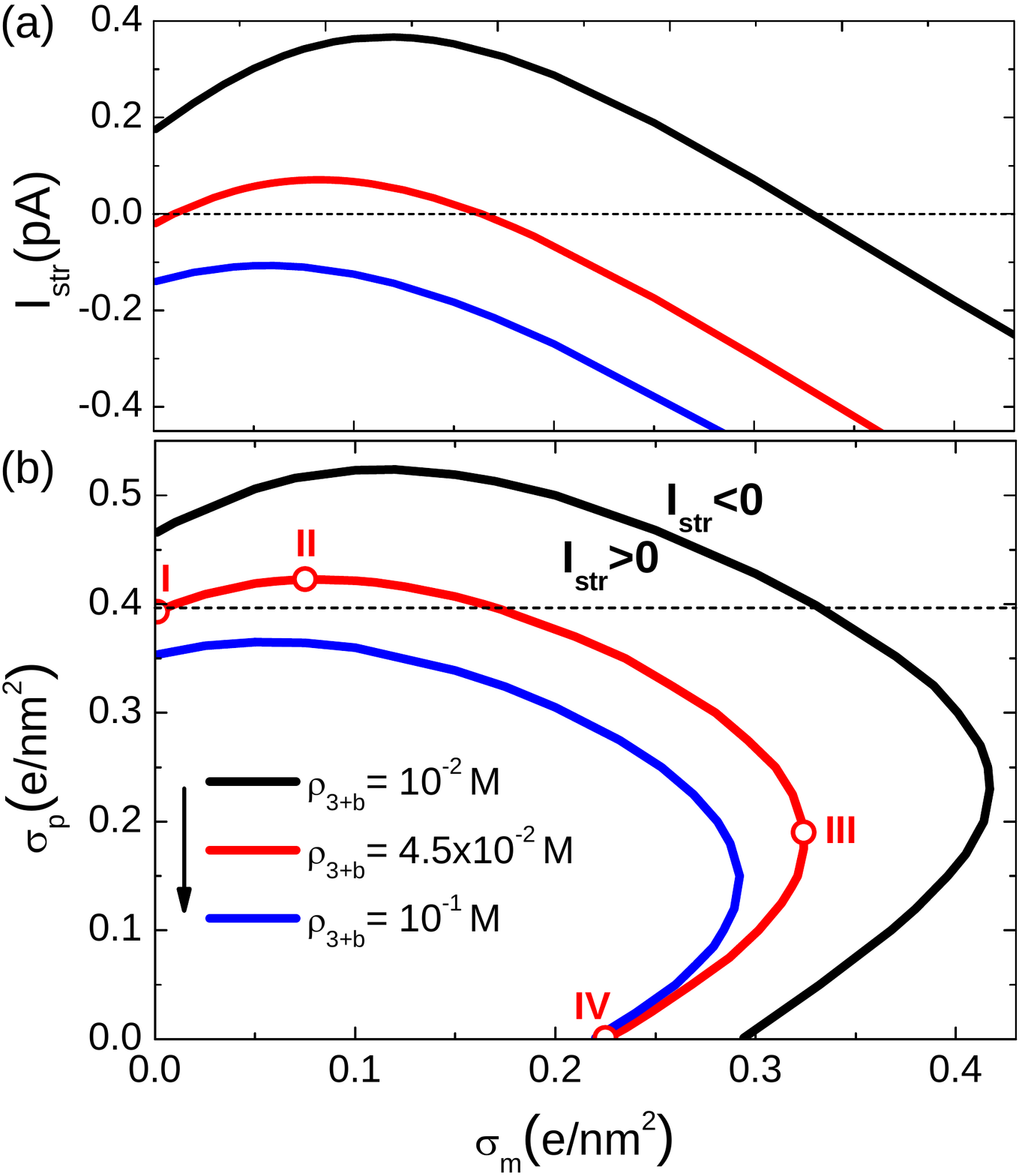}
\caption{(Color online) (a) Streaming current thorough the DNA-blocked pore ($\sigma_p=0.4$ $e/\mbox{nm}^2$) and (b) characteristic polyelectrolyte charges splitting the phase domains with positive and negative current against the surface charge density of the nanopore confining the  $\mbox{KCl}+\mbox{SpdCl}_3$ liquid. $\mbox{Spd}^{3+}$ densities corresponding to each curve in (a) and (b) are displayed in the legend of (b). The remaining parameters are the same as in Fig.~\ref{Fig2}.}
\label{Fig4}
\end{figure}

What is the effect of monovalent $\mbox{K}^+$ counterions on the streaming current inversion? In Fig.~\ref{Fig3}(b), we plot the critical $\sigma_p^*-\rho_{+b}^*$ curves separating the phase domains with positive and negative currents. One sees that at $\rho_{3+b}=0.01$ M (blue curve) and $\sigma_p=0.5$ $e/\mbox{nm}^2$, the increase of the $\mbox{K}^+$ density from $0.1$ M to $0.2$ M along the dashed blue line results in the inversion of the streaming current from negative to positive. In other words, $\mbox{K}^+$ ions drive the system back to the MF charge transport regime. We emphasize that such an effect has indeed been observed in pressure-driven nanofluidic experiments~\cite{Heyden2006}. In the phase diagram, one also sees that the higher is the polymer charge or the spermidine concentration (increased in the clockwise direction), the higher is the $\mbox{K}^+$ density needed to cancel the net current.  In order to better characterize the role of $\mbox{K}^+$ ions, in Fig.~\ref{Fig3}(c), we show that the increase of the $\mbox{K}^+$ density from $\rho_{+b}=0.1$ M to $0.2$ M weakens the positive branch of the electrostatic cumulative density. This in turn drives the hydrodynamic density from negative to positive. Thus, $\mbox{K}^+$ ions suppress the negative ionic current by cancelling the DNA charge reversal.

We finally characterize  the effect of the finite pore charge on the reversal of the streaming current. First of all, Fig.~\ref{Fig4}(a)  shows that the ion current rises with the wall charge up to $\sigma_m\approx0.1$ $e/\mbox{nm}^2$ where it reaches a peak and drops above this value. Then, one notes that depending on the $\mbox{Spd}^{3+}$ concentration, the non-monotonical shape of the streaming current curve may result in a single current inversion point ($\rho_{3+b}=0.01$ M),   two  inversion points ($\rho_{3+b}=0.045$ M), or no inversion point ($\rho_{3+b}=0.1$ M). To better understand the presence of multiple inversion points, we plotted in Fig.~\ref{Fig4}(b) the characteristic $\sigma_p-\sigma_m$ curves splitting the positive and negative current regions. In this figure, the current inversion points of Fig.~\ref{Fig4}(a) correspond to the intersections between the curves and the horizontal line marking the DNA charge. We split the black curve in Fig.~\ref{Fig4}(b)  into three segments. We found that on the segment I-II, the increase of the pore charge $\sigma_m$  has the main effect of bringing more $\mbox{K}^+$ ions  into the pore. This explains the increase of the pore conductivity ($\sigma_m\uparrow\mbox{I}_{\rm{str}}\uparrow$) in Fig.~\ref{Fig4}(a).  Furthermore, the branch II-III of the critical curve corresponds to the regime of strongly charged nanopores where correlations result in the reversal of the nanopore charge even in the absence of the polyelectrolyte. As a result, the attraction of $\mbox{Cl}^{-}$ ions to the charge inverted nanopore wall decreases the streaming current ($\sigma_m\uparrow\mbox{I}_{\rm{str}}\downarrow$) and switches the latter from positive to negative in Fig.~\ref{Fig4}(a). Fig.~\ref{Fig4}(b) also predicts that at the nanopore charges located on the interval III-IV, the ionic current inversion is expected to occur at two different polyelectrolyte charge densities. This complex dependence of the streaming current on the nanopore and polymer charge calls for experimental verification.

To conclude, we have predicted streaming current inversion induced by translocating polyelectrolytes in the presence of multivalent counterions. We have shown that at physiological charge densities, streaming current reversal upon ds-DNA  penetration takes place only if the multivalent charges in the electrolyte are trivalent or of higher valency. We have also found that ionic current inversion is cancelled by monovalent cations but favored by the nanopore charge. Our predictions can be easily verified by current pressure-driven translocation experiments. The proposed current inversion mechanism may also find applications in lab-on-a-chip technologies.

SB gratefully acknowledges support under the french ANR blanc grant ``Fluctuations in Structured Coulomb Fluids'' during the realization of the first part of the project. T. A-N. has been supported in part by the Academy of Finland through its CoE program COMP grant no. 251748 and by Aalto University's energy efficiency program EXPECTS.
\\


\begin{thebibliography}{99}
\bibitem {rem0} In our numerical calculations, the Stern layer is taken finite only if the corresponding surface carries a finite charge.
\end{thebibliography}
\end{document}